\documentclass[letterpaper]{JHEP3}
\usepackage{graphicx}
\usepackage{epsfig}
\usepackage{amssymb, amsmath}

\def \lf {\left (}
\def \rt {\right )}
\def \bra {\langle}
\def \ket {\rangle}

\title{Universal Correction to the Inflationary Vacuum}

\author{Brian~R.~Greene\footnote {\tt greene@phys.columbia.edu}
, Maulik~K.~Parikh\footnote {\tt mkp@phys.columbia.edu} , and Jan
Pieter van der Schaar\footnote {\tt jpschaar@phys.columbia.edu}
\\ Institute for Strings, Cosmology and Astroparticle Physics,
Department of Physics, Columbia University, New York, NY 10027 }

\abstract{The Bunch-Davies state appears precisely thermal to a
free-falling observer in de Sitter space. However, precise
thermality is unphysical because it violates energy conservation.
Instead, the true spectrum must take a certain different form, with
the Boltzmann factor $\exp(-\beta \omega_k)$ replaced by
$\exp(\Delta S)$, where $S$ is the entropy of the de Sitter horizon.
The deviation from precise thermality can be regarded as an
explicitly calculable correction to the Bunch-Davies state. This
correction is mandatory in that it relies only on energy
conservation. The modified Bunch-Davies state leads, in turn, to an
${\cal O} (H/M_p)^2$ modification of the primordial power spectrum
of inflationary perturbations, which we determine.}

\preprint{CU-TP-1136}

\begin{document}

\section{Introduction}

One of the most satisfying aspects of the inflationary paradigm is that
the enormous structures in the present universe actually arose
out of minuscule quantum fluctuations in the far past.
As the spacetime geometry of the universe
during inflation was, to excellent approximation, that of de Sitter space,
this means that the large-scale structure visible today is intimately
tied to the expectation value, $\bra \phi^2 \ket$, of quantum
fields in de Sitter space. Moreover, through observations of the
cosmic microwave background and through galaxy surveys, we are able to
indirectly probe the initial state of the universe.

Traditionally, the initial state in which the expectation values are
calculated is assumed to be the Bunch-Davies state, $|BD \ket$. This
state is chosen, firstly, because the corresponding Green's function
resembles that of the Poincar\'e-invariant vacuum at short distances
and, secondly, because (for a massive scalar field) the Bunch-Davies
state is invariant under the full de Sitter isometry
group. Furthermore, the expectation value of the inflaton in the
Bunch-Davies state, $\bra BD|\phi^2|BD \ket$, leads to the familiar,
and observationally corroborated, scale-invariant primordial power
spectrum of scalar density perturbations.

However, the assumption of the Bunch-Davies state as the initial
state has recently been questioned \cite{MB1, Niem,
EGKS1,EGKS2,EGKS3, Dan1, Dan2, Lowe, BGV}. Several plausible
alternatives for the initial state have been suggested based on the
anticipated behavior of quantum fields at Planckian energies
\cite{KN, BH, HS, NPC, ACT, Burgess1, BEFT1, BEFT2, Cremonini}.
Because the initial state is modified, these alternatives lead to
slightly different predictions for the primordial power spectrum. This
is an exciting possibility because the signatures of the modified
initial state may be within the realm of observational investigation
\cite{Brandenberger, BD, Hanestad, Martin, GSSS, Easther, Ash}.
Nevertheless, one drawback so far has been that the assumed behavior
of the fields at Planckian energies remains model-dependent.

In this paper, we take a different approach. We use the fact that
energy must be conserved. Energy conservation does not break down in
string theory and it would be surprising if we could not assume it
during inflation. This suggests a modification
of the Bunch-Davies state, as we now argue. It is well known that an
inertial observer in de Sitter space is immersed in a bath of de
Sitter radiation emanating from the de Sitter horizon \cite{GH}. The
spectrum of this radiation (not to be confused with the primordial
power spectrum) is, to first approximation, thermal. Indeed, it can
be shown that, in the Bunch-Davies state, the spectrum is {\em
precisely} the thermal Planck distribution characterized by the de
Sitter temperature.

In fact, however, the true spectrum cannot be precisely thermal.
Instead, as is known from parallel work on black hole radiation,
implementing energy conservation modifies the spectrum in a definite
way \cite{Kraus-Wil, Par-Wil}. One can regard the de Sitter
radiation, whose origin is quantum-mechanical, as emerging out of
virtual pairs of particles, one member of which has materialized by
tunneling across the horizon. The probability of emission of a
quantum of energy $\omega_k$ is then roughly
\begin{equation}
\Gamma_k \approx \exp(-\beta \omega_k) \; .
\end{equation}
That is, it is roughly thermal. However, it is not exactly thermal
because the horizon suffers a back-reaction when a particle is
emitted, as a consequence of energy conservation. In a derivation in
which energy conservation is taken into account \cite{Par1}, the
effect is that the thermal Boltzmann factor is replaced by $\exp(\Delta S)$:
\begin{equation}
\exp(-\beta \omega_k) \to \exp(\Delta S(\omega_k)) \; ,
\end{equation}
where $S$ is the Bekenstein-Hawking entropy of the de Sitter
horizon, and $\Delta S(\omega_k)$ is the change in the entropy when
a quantum of radiation with energy $\omega_k$ is emitted. Since the
Bunch-Davies state resulted in a perfectly thermal spectrum, we may
regard this new modified spectrum as a consequence of a small
modification of the initial state:
\begin{equation}
|BD \ket \to |BD' \ket \; .
\end{equation}
In turn, we are led to a robust and explicitly calculable modification
to the primordial power spectrum of inflationary perturbations:
\begin{equation}
\frac{2\pi^2}{k^3} P(k) = \int d^3x ~ e^{-i\vec{k} \cdot \vec{x}} ~
\bra BD'|\phi(\vec{x}) \phi(0) |BD' \ket \; .
\end{equation}
The modification to the primordial power spectrum thus arises
unavoidably by energy conservation via a correction to the
Bunch-Davies state.

This paper proceeds as follows. In section 2, we review the different vacuum
choices in de Sitter space. We also substantiate the following
interesting suggestion:
the Bunch-Davies state actually appears empty, rather than thermal, to a
hypothetical lightlike observer. In section 3, we state the results of the
tunneling derivation of de Sitter radiation. In section 4, we present a
formalism for obtaining the modification to the Bunch-Davies state
from the modification to the thermal spectrum. In section 5, we compute the
primordial power spectrum in this modified initial state. The result, a
correction of order $(H/M_p)^2$, while small, is universal; as such, it is a
model-independent signature of quantum gravity in the sky.

\section{Vacuum States in de Sitter Space}

A well known feature of quantum field theory, and one that makes
itself particularly manifest in curved spacetime backgrounds, is
that the very definition of a particle excitation and hence of a
thermal state depends sensitively on the choice of vacuum state.
In a general curved spacetime, there is no canonical or even preferred
vacuum state. However, if a spacetime admits isometries, and, in
particular, a timelike Killing vector field, then this provides a
natural means of partitioning modes into positive and negative
frequency categories and then, in line with the standard procedure in
Minkowski space, associating these with annihilation and creation
operators. A vacuum state can then be defined by imposing the
condition that the state be annihilated by all the annihilation
operators.

In the absence of an everywhere timelike Killing vector, there is no
natural choice of vacuum state. In that case, one can apply different
criteria to motivate particular choices of vacuum. A simple choice,
where possible, might be to restrict one's interest to a region of
spacetime for which a timelike Killing vector does exist, and to use
the corresponding vacuum state. Similarly, if the spacetime admits an
asymptotically Minkowskian region, another possibility is to use the
natural Poincar\'e vacuum in that region. Alternatively, one
could demand that the vacuum be annihilated by the generators of some
symmetry group. A vacuum state can also be deemed unphysical if it
fails to satisfy certain criteria. For example, if the expectation
value of the stress tensor diverges at a nonsingular point in
spacetime, such as at a horizon, one would consider this grounds to
reject the underlying vacuum state. In our case, we will require a
certain universal form for the particle spectrum detected by an
inertial observer, a form dictated by the conservation of energy.

In this section, we review some of the important vacua of de Sitter
space, as well as some affiliated coordinate systems.

\subsection*{The Bunch-Davies state}

The Bunch-Davies vacuum state, $|BD \ket$, is usually
defined using planar coordinates,
\begin{equation}
ds^2 = - dt^2 + e^{2Ht} dx^2 = \frac{1}{(H \eta)^2} \lf - d \eta^2 + d
\rho^2 + \rho^2 d \Omega^2 \rt \; ,
\end{equation}
where $\eta = - e^{-Ht}/H$ is conformal time. See Figure 1. The mode
solutions to the massless scalar field wave equation,
\begin{equation}
\label{deSittermodes}
u_{k}(\eta, \vec{x}) = N_k  \lf 1 + ik \eta \rt e^{-i k \eta + i
  \vec{k} \cdot \vec{x}} \; ,
\end{equation}
are termed positive frequency modes because they satisfy
\begin{equation}
\label{deSitterBoundary}
\frac{\partial}{\partial\eta}u_k(\eta,\vec{x}) = -iku_k(\eta, \vec{x})
\end{equation}
in the infinite past i.e. in the $\eta\rightarrow-\infty$ limit. An
oft-cited motivation for this boundary condition is that, in this
limit, the physical wavelength of any given mode is arbitrarily short
compared to the Hubble length, and hence any distinction between de
Sitter space and Minkowski space should be suppressed. The boundary
condition (\ref{deSitterBoundary}) is recognized as the standard Minkowski
space boundary condition, written in de Sitter space.

\FIGURE[r]{\epsfig{file=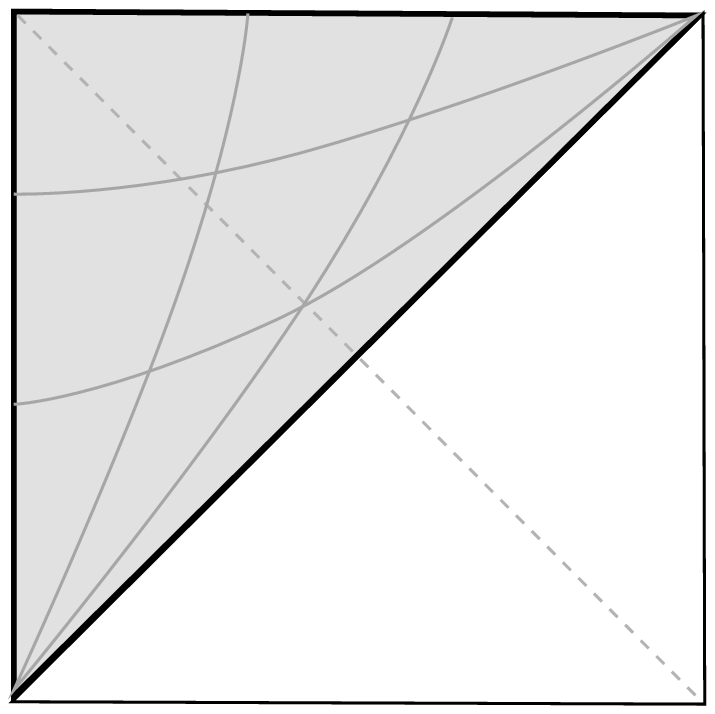, width = 6.2cm, height=6.2cm}
\caption{Penrose diagram for de Sitter space. Planar coordinates
cover
  the shaded region. The curves are sections of constant $\eta$ and
  constant $\rho$.}}

Another distinguishing property of the collection of modes
(\ref{deSittermodes})  is that, under an arbitrary $SO(4,1)$ de Sitter
transformation, the positive frequency modes each mix amongst
themselves, the negative frequency modes mix amongst themselves, but
these two classes of modes do not mix among each other. This
implies, in particular, that the Bunch-Davies vacuum state is de
Sitter invariant\footnote{Even though, formally, no de
Sitter-invariant vacuum exists for a massless scalar field
\cite{Allen}, by assuming a tiny but nonzero mass, we can ignore this
subtlety.}.

Even so, it is well known that the Bunch-Davies vacuum is not
uniquely de Sitter invariant; for massive scalar fields, there is a
one-parameter family of de Sitter-invariant vacua called
$\alpha$-vacua. These are related to the Bunch-Davies vacuum by
choosing the positive frequency modes to be
\begin{equation}
\label{alphavacua} u_k^{\alpha}(\eta, \vec{x}) = A \, u_k
(\eta,\vec{x}) + B \, u_{-k}^{*}(\eta,\vec{x}) \; ,
\end{equation}
with $A, B$ parameterized according to $A=\frac{1}{\surd(1-e^{\alpha
+ \alpha^*})}$ and $B = e^{\alpha}A$. Here, $\alpha$ is a complex
number labeling the $\alpha$-vacuum; the Bunch-Davies modes
correspond to $B = 0$ i.e. to $\alpha = - \infty$. A somewhat more
transparent expression of the $\alpha$ modes is to introduce the
antipodal map $(\eta, \vec{x})\rightarrow(-\eta, \vec{x}) \equiv
(\overline{x})$ in terms of which (\ref{alphavacua})
can be written
\begin{equation}
\label{alphavacuaantipodal} u_k^{\alpha}(x)= A \, u_k(x) + B \,
u_{-k}(\overline{x}) \; .
\end{equation}
Under an arbitrary de Sitter transformation, the positive and
negative frequency $u_k^{\alpha}$ don't mix, and hence an $\alpha$-vacuum
state, defined by the condition that it lies in the kernel of all
annhilation operators, is also de Sitter invariant.

\subsection*{The static state}
Another important state in de Sitter space is the static vacuum,
$| S \ket$. Consider de Sitter space in static coordinates:
\begin{equation}
\label{static}
ds^2 = -(1 - H^2 r^2) dt^2 + (1-H^2 r^2)^{-1} dr^2 + r^2 d \Omega^2 \; .
\end{equation}
See Figure 2. The time coordinate here is the proper time of a static
geodesic observer at the origin and $r=H^{-1}$ corresponds to the horizon
in de Sitter space.

\FIGURE[r]{\epsfig{file=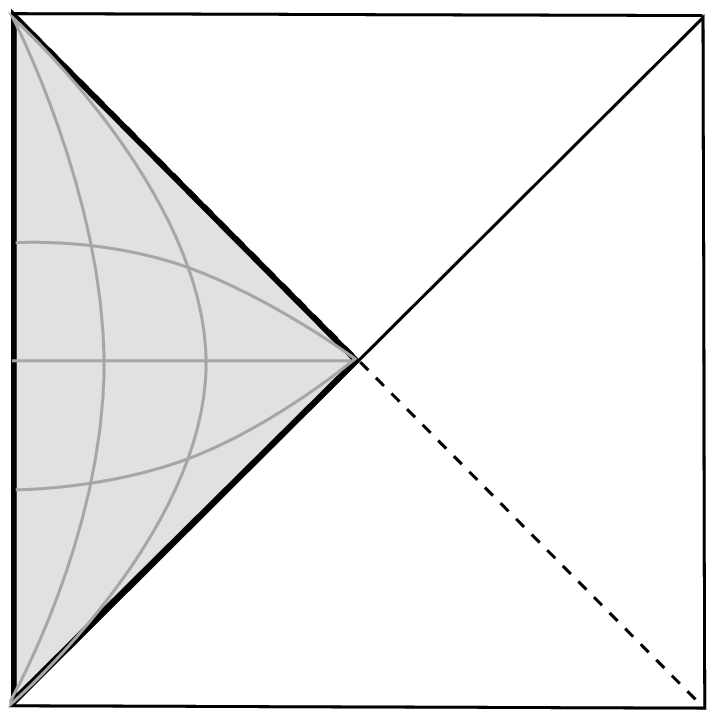, width = 6cm, height=6cm}
\caption{Penrose diagram for de Sitter space. Static
    coordinates cover  the shaded region. The curves are sections of
    constant $t$ and constant $r$.}}

Using this time coordinate to define positive and
negative frequencies leads to the static vacuum, so called because the
generator of time translations, $\partial_t$, is a Killing vector.
The physical significance
of the static vacuum is that the particular geodesic observer
resting at $r = 0$ sees it as empty: $b_k |S \ket = 0$ for all $k$.
Of course there is nothing special about this observer, and, indeed,
there exists a corresponding vacuum state for each of the infinitely
many freely falling observers that crisscross de Sitter space. The
static vacuum is invariant under $SO(3) \times R$, i.e. under
rotations and time translations, but not under the entire de Sitter
isometry group, $SO(4,1)$. What seems to disqualify the static vacuum
as a consistent state in de Sitter space is that the expectation
value of the energy-momentum tensor blows up as one approaches the de
Sitter horizon \cite{BirrellDavies}.

\subsection*{Relation between the Bunch-Davies state and the static state}
It can be shown that the Bunch-Davies state appears precisely thermal
to the static observer with temperature
\begin{equation}
T = \frac{H}{2 \pi} \; .
\end{equation}
The easiest way to see this is to consider the Bunch-Davies Green's
function, $G(x,x')$. One sets $x(\tau)$ and $x'(\tau')$ to lie along a
timelike geodesic with proper time $\tau$; the observer moving along
such a geodesic registers a thermal spectrum if two conditions hold.
First, the Green's function needs to be periodic under a shift $\tau \to
\tau + \beta$ (where $\beta = T^{-1}$), and second, the singularities
of the two-point Green's function must allow for contour integrals
that are consistent with detailed balance. In this way, it can be
shown that the Bunch-Davies state appears thermal to the static
observer, and, indeed, by de Sitter symmetry, to all timelike
geodesic observers.

While the Bunch-Davies vacuum is a thermal state, the same is not
true of any of the $\alpha$-vacua for $\alpha \neq -\infty$. For $B
\neq 0$, the combination in (\ref{alphavacuaantipodal}) introduces
additional singularities for the two-point Green function in the
complex plane, which in turn spoils the thermality of the state. The
$\alpha$-vacua deviations from a thermal spectrum can be calculated
by properly accounting for the extra singularities \cite{Burgess,
BMS, KKLSS}, or by using a Bogolubov approach as illustrated in
Appendix A.

Perhaps, then, the best way to characterize the Bunch-Davies vacuum
is its being the unique de Sitter-invariant state that appears
thermal to a freely falling observer. Indeed, this is an accurate
characterization of the Bunch-Davies state, but what is less widely
appreciated is that there are an infinity of other states that fail
to meet this very same set of criteria in the most mild of ways.

Here is what we mean. Consider a freely falling observer in de
Sitter space, using static coordinates. Since
$\partial_t$ is a manifest Killing vector in these
coordinates, we know that
\begin{equation}
\label{tgeodesic} u_0 = -(1-H^2 r^2)\frac{dt_s}{d\tau}
\end{equation}
is conserved along geodesics. Here $\tau$ is the proper time and $u^\mu$
are the components of the velocity four-vector (which should not be
confused with mode functions). Let's call the value of this conserved
quantity $-E$. Since $u^2 = -1$ we also have
\begin{equation}
\label{rgeodesic}
\lf \frac{dr}{d\tau} \rt^{\! 2} = E^2 - (1-H^2 r^2) \; .
\end{equation}
Solving (\ref{tgeodesic}) and (\ref{rgeodesic}) subject to the
boundary conditions $t_s(\tau = 0) = 0, r(\tau = 0) = 0$ we find
\begin{eqnarray} \label{geosolution}
r(\tau) & = & \frac{1}{H} \sqrt{(E^2 -1)} \sinh (H \tau) \nonumber \\
t_s(\tau) & = & \frac{1}{H} \tanh^{-1} ( E \tanh (H \tau)) \; .
\end{eqnarray}
When $E=1$, we find that $r(\tau)=0$ and $t_s(\tau)=\tau$, which
parameterizes the worldline of the static observer who stays put at $r=0$.
In contrast, notice that for any $E \neq 1$, $t_s$ is invariant under
$\tau \rightarrow \tau + \pi i H^{-1}$ while $r$ is invariant under $\tau
\rightarrow \tau + 2 \pi i H^{-1}$.

Consider now the two-point Green's function for a scalar field, making
use of the static vacuum, i.e. the zero-particle state as seen by a
given $E = 1$ observer. Since this two-point Green's function when
evaluated along a geodesic is a function of $(t(\tau),r(\tau))$, we
see that the Green's function is invariant under $\tau \rightarrow
\tau + 2 \pi i H^{-1}$. Moreover, since in this analysis we have not
modified the static patch Green function -- we have only evaluated it
along a particular curve -- we can be sure that no new singularities
have been introduced (in contrast to the case with $\alpha$-vacua).
Thus the periodicity of the Green's function implies that any freely
falling observer (except for the sole such observer with
$E=1$) measures a thermal spectrum of particle excitations.

In other words, the static state as defined by the static ($ E =
1$) observer, appears to be a thermal state for all other freely
falling observers. The only freely falling observer who
does not see a thermal spectrum is the static observer at $r =0$, for
whom this is the vacuum. Of course, nothing is special about this
observer in our analysis; we can do the same analysis for every other
freely falling observer. Doing so, we conclude that any freely falling
observer can define a vacuum state with respect to which his or her
``Unruh'' particle detector will fail to detect any particles. The
Unruh detectors of {\em all} other freely falling observers, however,
will measure a thermal spectrum with the same temperature.

There is an intuitive way to understand this. Recall that, for black
holes, the state that seems empty to the infalling observer looks
thermal to the outside observer. Since the derivations are identical,
this implies that, in de Sitter space, the state that seems empty to
an observer who falls through another observer's horizon will seem
thermal to the other observer. It follows by de Sitter symmetry that
the static vacuum of any observer looks thermal to all other observers.

{}From the perspective of thermality, therefore, the distinction
between the Bunch-Davies vacuum and the static patch vacua just
described is simply this: All freely falling observers see the
Bunch-Davies vacuum as a thermal state. All but one freely falling
observer sees a given static patch vacuum state as thermal. That one
special observer, of course, is the observer with respect to which
the static vacuum is defined.

Heuristically, this suggests thinking about the Bunch-Davies vacuum
state in the following way. All observers moving along timelike
geodesics see the Bunch-Davies vacuum state as thermal. As above, all
but one observer sees a given static patch vacuum state as thermal. If
this one special observer should actually be moving along a lightlike
trajectory, then the odd man out -- the one observer who does not see
the static patch vacuum state as thermal -- would not be among the
timelike observers. This suggests that one might think of the
Bunch-Davies vacuum as that state which appears empty to an observer
moving on a lightlike trajectory, but which appears thermal to all
observers moving on timelike trajectories. From this perspective, the
Bunch-Davies vacuum and the static patch vacua are all part of a
single class of vacuum states, each appearing empty to one observer
and thermal to everyone else.

One clue that this heuristic picture is correct is that the light ray
emanating from $r=0$ in the limit $\eta \rightarrow -\infty$ actually
travels along the $\eta \rightarrow -\infty$ time slice.
But as $\eta \to -\infty$, the Bunch-Davies state becomes the
ordinary vacuum of Minkowski space. Now, it is known that, in
Minkowski space, the  Poincar\'e-invariant vacuum is the same as the
light-cone vacuum i.e. the vacuum with respect to the Hamiltonian
generating lightlike translations. Thus the Bunch-Davies state is the
same as the vacuum state for a lightlike observer.
To make this more explicit, we note that the Bunch-Davies
modes become plane waves in the far past. Picking an arbitrary spatial
direction $x$, one expands the field operator in the $\eta \to
-\infty$ limit as
\begin{eqnarray}
\phi(\eta,x,\vec{x}) = \int_0^\infty \frac{d k_x}{\sqrt{2 \pi}}
\int_{-\infty}^{+\infty} \frac{d^2 k_T}{2 \pi}
\frac{1}{\sqrt{2\omega_k}} && \! \! \! \! \! \Big\{ \lf
a_{k_x,\vec{k}_T} e^{-i \omega_k \eta + i k_x x + i \vec{k}_T \cdot \vec{x}_T}
+ {\rm h.c.} \rt \nonumber \\
&& \! \! \! \! \! + \lf  a_{-k_x, \vec{k}_T} e^{-i \omega_k \eta - i k_x x + i
\vec{k}_T \cdot \vec{x}_T} + {\rm h. c.} \rt \Big\} \; ,
\end{eqnarray}
where $x_T$ and $k_T$ are the transverse position and wave number.
The Bunch-Davies state is defined by $a_{k_x, \vec{k}_T} |BD \ket =
a_{-k_x, \vec{k}_T} |BD \ket = 0$. But now these modes can be
written in light cone coordinates $x^{\pm} = \eta \pm x$:
\begin{eqnarray}
\phi(x^{\pm},\vec{x}) = \int_0^\infty \frac{d k_x}{\sqrt{2 \pi}}
\int_{-\infty}^{+\infty} \frac{d^2 k_T}{2 \pi}
\frac{1}{\sqrt{2\omega_k}} && \! \! \! \! \! \Big\{ \lf a_{k_x,
  \vec{k}_T} e^{-i [(\omega_k - k_x) x^+ + (\omega_k + k_x) x^-] + i
  \vec{k}_T \cdot \vec{x}_T} + {\rm h.c.} \rt \nonumber \\
&& \! \! \! \! \! \! \! \! \! \! \! \! \! \! \! \! \! \! \! \! \! \!
\! \! \!
+ \lf a_{-k_x, \vec{k}_T} e^{-i
  [(\omega_k + k_x) x^+ + (\omega_k - k_x) x^-] + i \vec{k}_T \cdot
  \vec{x}_T} + {\rm   h. c.} \rt \Big\} \; .
\end{eqnarray}
Since $(\omega_k \pm k_x) \geq 0$,
the positive frequency modes defining the Bunch-Davies
vacuum also correspond to the positive frequency modes defining the
light-cone vacuum. Accordingly, since one of the light-cone
coordinates $x^\pm$ corresponds to the affine parameter
along a lightlike trajectory, the Bunch-Davies vacuum, being
equivalent to the light-cone vacuum, also appears empty with respect
to a fiducial observer traveling along a lightlike trajectory.

Formally, we can regard the transformation that takes a
timelike observer into a lightlike observer as an infinite
boost. Infinite boosts are not contained in the de Sitter group, which
is noncompact, but one can include them by extending the de Sitter
group to include its one-point compactification. Under the de Sitter
group, the Bunch-Davies state is special in that it is thermal to all
geodesic observers. But under the compactified de Sitter group the
Bunch-Davies state is merely in the same class of states as the static
vacua.

\subsection*{Painlev\'e coordinates}

Painlev\'e coordinates \cite{Par1} are a particularly useful
coordinate system. The line element is
\begin{equation}
\label{Painleve}
ds^2 = -\left( 1-H^2 r^2 \right) \,dt^2 - 2Hr\, dr \,
dt + \, dr^2 + r^2 \, d\Omega^2 \; .
\end{equation}
\FIGURE[r]{\epsfig{file=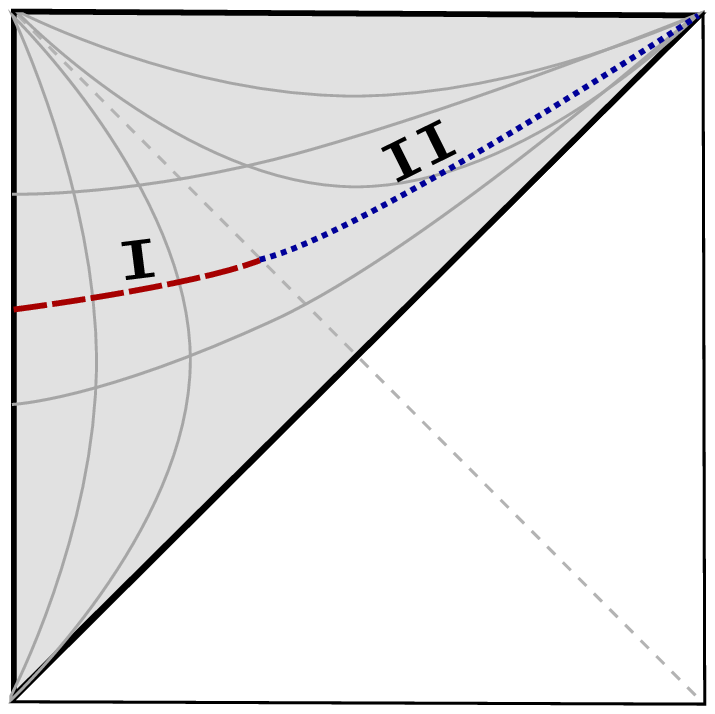, width = 6cm, height=6cm}
\caption{Penrose diagram for de Sitter space. Painlev\'e coordinates cover
  the shaded region. The curves are sections of constant $t$ and
  constant $r$. Also shown is a constant-time slice split into
  two.}}
These coordinates resemble static coordinates, but have the advantage
that the components of the metric and its inverse are finite at the
horizon. Moreover, constant-time slices correspond to flat Euclidean
space. In fact, the time slices are equivalent to the constant-time slices
using planar coordinates. See Figure 3. Indeed, Painlev\'e coordinates
are a hybrid of static and planar coordinates, sharing the radial
coordinate of the former but the time coordinate of the latter:
\begin{eqnarray}
\label{coordtransforms}
t_{\rm Painleve} & = t_{\rm planar} & = t_{\rm static} + \frac{1}{2H}
\ln (1 - H^2 r^2) \nonumber \\
r_{\rm Painleve} & = r_{\rm static} & = \rho_{\rm planar} ~ e^{Ht} \; .
\end{eqnarray}
Physically, Painlev\'e time is the proper time of comoving observers,
as Painlev\'e time is the same as planar time.

What is the vacuum associated with Painlev\'e time? Because of the
off-diagonal metric component, the time translation vector $\partial_t$
is different from the normal vector, $n$, to surfaces of constant
$t$. Should one consider modes to be positive frequency if they have
positive eigenvalues with respect to $\partial_t$ or with respect to $n$?
The answer follows immediately from the Klein-Gordon inner product,
\begin{equation}
\label{KG}
(\phi_1, \phi_2) = i \int_{\Sigma} d \Sigma^a (\phi_1^*
 \overleftrightarrow{\partial_a} \phi_2) \; .
\end{equation}
In this expression, $d \Sigma^a = n^a d \Sigma$ is the oriented volume
element of a spatial slice, $\Sigma$. The inner product defines
positive norm modes, and, since these are paired with annihilation
operators, also defines the vacuum state. As the above expression
shows, positive frequency should be defined with respect to the normal
vector, $n$, not $\partial_t$. But Painlev\'e time and planar time are
the same, and hence, so are their time-slices. It follows that the
Painlev\'e vacuum is precisely the Bunch-Davies state. This is
important because the derivation of the approximately thermal spectrum
in the next section will imply a back-reaction modification to the
Painlev\'e state. So it must mean that the Bunch-Davies state is
modified.

\section{Deviations from Thermality}

In inflationary cosmology, the initial state of the universe is
usually taken to be the Bunch-Davies state. As explained above, the
motivation for this choice of state is two-fold: the state is de
Sitter invariant and the corresponding Green's function has a
short-distance behavior appropriate to Minkowski space. Now, the
Bunch-Davies state is a thermal state; an
Unruh detector for a free-falling observer registers a
precisely Planckian spectrum with temperature $H/2 \pi$.
However, there exist quite general reasons why the initial state cannot
be strictly thermal; the Bunch-Davies state is not physical
because it violates energy conservation. To see this,
consider the closely analogous background of a Schwarzschild black hole.

At first sight, the spectrum of a black hole's Hawking radiation is
also Planckian. Yet, for a variety of reasons, the true spectrum actually
departs from pure thermality. First, a purely thermal spectrum would have a
tail extending out to infinitely high energies. But a black hole
obviously cannot emit a quantum with more energy than its own mass. Thus
there is a high-energy cut-off to the spectrum. Moreover, even at
energies significantly lower than this cut-off, the spectrum must deviate
from strict thermality. This is because the ``temperature'' of a black
hole is a function of the mass -- but the mass is different before and
after emission.

Parallel arguments apply to de Sitter space. De Sitter space has a
horizon of radius $H^{-1}$. This puts an upper bound on the energy
of any quantum in de Sitter space: it cannot exceed the mass of the
``Nariai black hole,'' the largest black hole that can fit inside the
de Sitter horizon. Moreover, as we will see, when the de Sitter
horizon emits a quantum of energy, it shrinks, much as a black horizon
does. The horizon temperature is thus affected by the emission of the
quantum so that the spectrum has small deviations from thermality even
at lower energies. This effect become more prominent at higher energies.

Finally, there is also a more fundamental reason why, in a
gravitational setting, one does not have a well-defined notion of
temperature. Recall that to define a temperature for a system, one
formally places the system in contact with an infinite reservoir. But,
in the presence of gravity, no such reservoir exists: any system
larger than the Jeans length will collapse. One should therefore more
properly consider the microcanonical ensemble -- in which energy is
fixed -- rather than the canonical ensemble (for which the temperature
is fixed).

The above arguments indicate that the Bunch-Davies state, by
virtue of being precisely thermal, cannot be quite the right
state. Rather, the correct state is the one
that yields the spectrum appropriate to the microcanonical
ensemble. Once we have this spectrum, we will, in
the next section, determine the state that gives rise to it; we will then
calculate the modification to the power spectrum of primordial density
perturbations.

Fortunately, there already exists a treatment of de Sitter radiation
\cite{Par1} in a microcanonical framework. In this picture, de
Sitter radiation arises because a virtual pair is created on the
other side of the horizon. The positive-energy member of the pair is
forbidden outside the horizon, because, in static coordinates, the
$t-t$ component of the metric has the opposite sign there. This
positive-energy particle then tunnels across the horizon towards the
observer where, because of the flip in the sign of the metric, it
becomes classically allowed. This picture of radiation as a
tunneling phenomenon naturally incorporates energy conservation
because the barrier across which the particle tunnels is determined
by the change in the horizon radius. One finds \cite{Par1,Med} that
the probability of emission is
\begin{equation}
\Gamma \sim \exp (\Delta S) = \exp \left [ \frac{\pi}{G} (r_f^2 - r_i^2)
\right ] \; , \label{rate}
\end{equation}
where $\Delta S$ is the change in the entropy of the horizon, and
$r_i$ and $r_f$ are the initial and final positions of the horizon.
As we will see, at low energies (\ref{rate}) approximates a thermal
Boltzmann factor. We note that the form of the expression, as
anticipated in \cite{Peresko}, is exactly the same as for black
holes \cite{Par-Wil,Par2}. Another appealing aspect of this result
in that context is that it is consistent with an underlying unitary
quantum theory \cite{Par3}.\footnote{Were there a description of de
Sitter radiation in a unitary theory of quantum gravity, one would
expect the emission rate to be given by the square of the amplitude
times the phase space factor. The latter is given by summing over
the $e^{S_{\rm final}}$ final states and averaging over the
$e^{S_{\rm initial}}$ initial states. Thus we would expect
\begin{equation}
\Gamma = |{\rm amplitude}|^2 \times (\mbox{phase space factor}) \sim
       \frac{e^{S_{\rm final}}}{e^{S_{\rm initial}}} = \exp(\Delta S) \; ,
\end{equation}
in accord with our expression.}

To see that $r_i$ differs from $r_f$, consider
a single particle emitted by the de Sitter horizon. Before
emission, the spacetime is empty and is described by the line element
of de Sitter space, (\ref{static}).
Thus the initial horizon radius is simply
\begin{equation}
r_i = H^{-1} \; .
\end{equation}
After emission, the spacetime is not pure de Sitter space because it
contains a quantum of energy $\omega$. For simplicity, consider emission
in the s-wave so that the particle is really a spherical shell. The
spherically-symmetric spacetime with energy $\omega$
is Schwarzschild-de Sitter space. This has the line element
\begin{equation}
\label{SdS}
ds^2 = -(1 - H^2 r^2 - 2G \omega /r) dt^2 + (1- H^2 r^2 - 2G
\omega/r)^{-1} dr^2 + r^2 d \Omega^2 \; .
\end{equation}
The final location of the horizon is determined by setting $g^{rr}$ to
zero and solving the resulting cubic equation for its largest root. We
find
\begin{equation}
r_f = \frac{2}{H \sqrt{3}} \cos \lf \frac{1}{3} \arctan \lf
 \frac{\sqrt{\frac{1}{27} - (G \omega H)^2}}{-G \omega H} \rt \rt \; ,
\end{equation}
where the arctan takes values between $\pi/2$ and $\pi$. Equation (\ref{rate})
then implies an emission rate of
\begin{equation}
\Gamma \sim \exp \left [ \frac{\pi}{H^2 G} \lf \frac{4}{3} \cos^2 \lf
\frac{1}{3} \arctan \lf \frac{\sqrt{\frac{1}{27} - (G
    \omega H)^2}}{-G \omega H}
  \rt \rt - 1 \rt \right ] \; .
\end{equation}
This expression can be rendered more tractable by considering the low-energy
limit $G \omega H \ll 1$. Then
\begin{equation}
\Gamma \approx \exp \left [ - \frac{2 \pi}{H} \omega \lf 1 +
  \frac{\omega H}{8 \pi M_p^2} \rt \right ] \; , \label{lowenergy}
\end{equation}
where we have expressed Newton's constant in terms of the Planck
mass: $M_p^{-2} = 8 \pi G$. If we neglect the $\omega H/(8 \pi M_p^2)$
correction, we find a probability that is precisely the Boltzmann
factor for emission at temperature $H/2 \pi$. Had there been no
correction term, elementary statistical mechanics would then have
implied a spectrum with Planckian occupation numbers, precisely the
spectrum detected by a free-falling Unruh detector in the Bunch-Davies
state.

The correction term can be regarded as a consequence of
self-gravitation or, equivalently, of back-reaction or energy
conservation \cite{Kraus-Wil,Par2}.
In the next section, we will infer the corrected initial state that
corresponds to the corrected spectrum. Note that we can already
anticipate a modification to the primordial power spectrum. From
(\ref{lowenergy}), we can attribute an effective temperature
\begin{equation}
T_{\rm eff} \approx \frac{H}{2 \pi} \lf 1 - \frac{\omega H}
{8 \pi M_p^2} \rt \; ,
\end{equation}
so that, at typical energies, $\omega \sim H$, we have a modification of
order $(H/M_p)^2$, whereas at Planckian energies we find a
modification of order $H/M_p$.

\section{Determining the Correction to the Bunch-Davies State}

We will now interpret the corrections to the thermal spectrum in
terms of a small modification of the initial state. Let us first
explain the setup. In canonical language, the probability for
pair creation of quanta of energy $\omega_k$ is
\begin{equation} \label{defrate}
\Gamma_k = \left | \frac{\bra {\rm out} | \, b_k \,
b_{-k} \, | {\rm in} \ket}{\bra {\rm out} | {\rm in}
\ket} \right|^2 \; ,
\end{equation}
where the operators $b_k$ annihilate the out-vacuum, i.e. $b_k |
{\rm out} \ket=0$. We need to determine what $|{\rm in} \ket$ and
$|{\rm out} \ket$ refer to in the tunneling calculation. Now, that
calculation was performed using Painlev\'e coordinates\footnote{See
\cite{Kraus-Wil} for earlier work incorporating back-reaction using
a Hamiltonian formalism.} \cite{Par1, Med}. The tunneling
probability is the probability to go from empty de Sitter space to a
spacetime containing a pair of particles, one on either side of the
horizon, only one of which is detected as de Sitter radiation by the
observer at $r = 0$. This suggests that the out-state should be
related to the static vacuum $|S \ket$, as defined by an $r=0$
observer. But, more precisely, because Painlev\'e coordinates cover
the full planar patch, the out-state actually splits (see Figure 3)
into a tensor product:
\begin{equation}
\label{outstate}
| {\rm out} \ket = | {\rm I} \ket \otimes | {\rm II} \ket \; ,
\end{equation}
with the state in the static region ${\rm I}$ corresponding to the
static vacuum, $| {\rm I} \ket \sim | {\rm S} \ket$. That $|{\rm I}
\ket$ corresponds to the static vacuum can be seen in a couple of
ways. First, it is suggested by the near-thermal result of the path
integral. And second, in region ${\rm I}$, the timelike Killing vectors
$\partial_{t_{\rm static}}$ and $\partial_{t_{\rm Painleve}}$ agree,
by (\ref{coordtransforms})\footnote{
In principle one should be able to determine the structure
of the $|{\rm out} \ket$-state unambiguously from the path integral
derivation; it would be nice to show this formally.}.
Then since the quanta in the out-state have to be created in pairs,
and because only one member of the pair ends up on each side of the
horizon, only one of the creation operators acts on the static
vacuum. The other one acts on ${\rm II}$, on the other side of the
horizon. That is, the amplitude is for the transition $|{\rm in} \ket
\to b^{(I)\dagger}_k b^{(II)\dagger}_{-k} |{\rm out} \ket$. Thus the
probability of creating two particles in the out-state is the same as
the probability of detecting one particle in the static vacuum.

Furthermore, the in-state is, to first approximation, just the
Painlev\'e vacuum. We argued in section 2 that that was just the
Bunch-Davies state, $|BD \ket$. However, were the in-state really the
Bunch-Davies state, we would have obtained a precisely thermal spectrum. The
not-quite thermal spectrum suggests instead that the in-state is
a slightly modified Bunch-Davies state, $|BD' \ket$. Indeed, the
effective geometry seen by a self-gravitating shell is not pure de
Sitter space, but rather Schwarzschild-de Sitter space, (\ref{SdS}). This has a
slightly different Painlev\'e metric and is therefore associated with
a slightly different state.

In summary, we have three states: $|\rm out \ket$, $|BD \ket$, and
$|BD' \ket$. Moreover,
\begin{eqnarray}
\label{BD} \left | \frac{\bra {\rm out} | \, b_k \, b_{-k} \, | {\rm BD}
\ket}{\bra {\rm out} | {\rm BD}
\ket} \right|^2 & = & e^{-\beta \omega_k}  \nonumber \\
\left | \frac{\bra {\rm out} | \, b_k \,
  b_{-k} \, | {\rm BD'} \ket}{\bra {\rm out} | {\rm BD'} \ket} \right|^2
&=& e^{\Delta S(\omega_k)} \; ,
\end{eqnarray}
where the first expectation value is the standard thermal expression, and
the second expectation value was derived through the tunneling
calculation. Let us now use these expressions to determine the precise
relationship between the Bunch-Davies state and the modified state in
terms of Bogolubov coefficients.

\subsection{A general Bogolubov approach}

Consider a real quantum scalar field in the planar patch of de Sitter space.
Expand the scalar field in a complete set of orthonormal
momentum eigenmodes:
\begin{equation}
\label{spat-exp} \phi(x) = \sum_k \, \phi_k(x) \; ,
\end{equation}
with $k\equiv |\vec{k}|$. The time coordinate is assumed to be
Painlev\'e time (\ref{Painleve}), which itself is equivalent
to planar time. Define three different expansions of
the same quantum field:
\begin{eqnarray}
\label{1b}
\phi_k(x) &=& a_k ~ u_k(x) + a_{-k}^\dagger ~ u^*_{-k}(x) \nonumber \\
&=& {a'}_k ~ {u'}_k(x) + {a'}_{-k}^\dagger ~ {u'}^*_{-k}(x) \nonumber
\\ &=& b_k ~ v_k(x) + b_{-k}^\dagger ~ v^*_{-k}(x) \; .
\end{eqnarray}
Also define three different vacuum states, each corresponding to an
``empty'' state in terms of the associated annihilation operators:
\begin{equation}
\label{1c} a_k \, | BD \ket = 0 \; , \quad {a'}_k | BD' \ket =0 \; ,
\quad b_k  | {\rm out} \ket =0 \; .
\end{equation}
We have in mind that these three states are the Bunch-Davies state,
the modified Bunch-Davies state, and the out-state. As in
(\ref{outstate}), the out-state can be written as a tensor product of
states in regions ${\rm I}$ and ${\rm II}$; the vacuum state in region
${\rm I}$ is just the static vacuum.

The particle content of our three states will generally be nonzero
with respect to each other; Bogolubov transformations mix up the
positive and negative frequency modes. Expressing the linear
decomposition for the mode functions $v_k(x)$ in terms of $u_k(x)$
or $u'_k(x)$ we have
\begin{eqnarray}
\label{2c} v_k(x) &=& \sum_j \left[ \tilde{\alpha}_{jk} ~ u_j(x) +
  \tilde{\beta}_{jk} ~ u_{-j}(x)^* \right] \nonumber \\  \label{2d}
&=& \sum_i \left[ \tilde{\alpha}'_{ik} ~ u'_i(x) +
\tilde{\beta}'_{ik} ~ u'_{-i}(x)^* \right] \; .
\end{eqnarray}
Similarly, the different creation and annihilation operators are related via
\begin{eqnarray}
\label{2a}
b_k &=& \sum_j \left[ \tilde{\alpha}^*_{jk} ~ a_j - \tilde{\beta}^*_{jk} ~
  a_{-j}^\dagger \right] \nonumber \\ \label{2b}
&=& \sum_i \left[ \tilde{\alpha}'^*_{ik} ~ {a'}_i -
\tilde{\beta}'^*_{ik} ~
  {a'}_{-i}^\dagger \right] \; .
\end{eqnarray}
Proper normalization of the scalar field requires that
\begin{eqnarray}
\label{2e} \sum_k \left( \tilde{\alpha}_{ik} \,
\tilde{\alpha}^*_{jk} - \tilde{\beta}_{ik} \, \tilde{\beta}^*_{jk}
\right) = \sum_k \left( \tilde{\alpha}'_{ik} \,
\tilde{\alpha}'^*_{jk} - \tilde{\beta}'_{ik} \,
\tilde{\beta}'^*_{jk}  \right) &=& \delta_{ij} \, \nonumber \\
\sum_k \left( \tilde{\alpha}_{ik} \, \tilde{\beta}_{jk} -
\tilde{\beta}_{ik} \, \tilde{\alpha}_{jk}  \right) = \sum_k \left(
\tilde{\alpha}'_{ik} \, \tilde{\beta}'_{jk} - \tilde{\beta}'_{ik} \,
\tilde{\alpha}'_{jk}  \right) &=& 0 \; .
\end{eqnarray}
A general Bogolubov transformation, viewed as a matrix, can have
nonvanishing off-diagonal components. However, for our present
situation, because the spatial mode-functions are defined on the same
spatial slice for all three states (\ref{spat-exp}), we can choose an
orthonormal basis such that the Bogolubov coefficients are diagonal in $k$:
\begin{eqnarray}
\tilde{\alpha}_{kl} &\equiv& \tilde{\alpha}_k \delta_{kl} \quad ,
\quad \tilde{\beta}_{kj} \equiv \tilde{\beta}_k \delta_{kj} \quad ,
\quad |\tilde{\alpha}_k|^2 - |\tilde{\beta}_k|^2 = 1
\nonumber \\
\label{ucon} \tilde{\alpha}'_{kl} &\equiv& \tilde{\alpha}'_k
\delta_{kl} \quad , \quad \tilde{\beta}'_{kj} \equiv
\tilde{\beta}'_k \delta_{kj} \quad , \quad |\tilde{\alpha}'_k|^2 - |
\tilde{\beta}'_k|^2 =1 \; .
\end{eqnarray}

We are interested in solving the following problem: Suppose we know
the Bogolubov coefficients $\tilde{\alpha}_k$, $\tilde{\beta}_k$ and
$\tilde{\alpha}'_k$, $\tilde{\beta}'_k$, relating the $| BD \ket$
and $| BD' \ket$ states to the $\left |{\rm out} \right
>$ state. How do we then determine the Bogolubov coefficients between
the $| BD \ket$ and $| BD' \ket$ states? In other
words, we want to express the Bogolubov coefficients $\alpha_k$ and
$\beta_k$, defined by
\begin{equation} \label{4a}
{a'}_k = \alpha^*_k ~ a_k - \beta^*_k ~ a_{-k}^\dagger
\end{equation}
in terms of the coefficients $\tilde{\alpha}_k$, $\tilde{\beta}_k$
and $\tilde{\alpha}'_k$ and $\tilde{\beta}'_k$. Since the final
result will be important for the rest of the paper, let us be as
explicit as possible. Multiplying (\ref{2b}), by
$\tilde{\alpha}_k^{\prime*}$ and its complex conjugate by
$\tilde{\beta}'_k$, and subtracting, we see that
\begin{equation} \label{5a}
{a'}_k = \tilde{\alpha}'_k ~ b_k + \tilde{\beta}'^*_k ~
b_{-k}^\dagger \; .
\end{equation}
Using (\ref{2a}), we find
\begin{equation} \label{5b}
{a'}_k = \lf \tilde{\alpha}'_k \tilde{\alpha}^*_k -
\tilde{\beta}'^*_k \tilde{\beta}_k \rt ~ a_k + \lf -
\tilde{\alpha}'_k \tilde{\beta}^*_k + \tilde{\beta}'^*_k
\tilde{\alpha}_k \rt ~ a_{-k}^\dagger \; ,
\end{equation}
so that, by (\ref{4a}),
\begin{eqnarray}
\label{6} \alpha^*_k &=& \tilde{\alpha}'_k \tilde{\alpha}^*_k -
\tilde{\beta}'^*_k \tilde{\beta}_k  \nonumber \\
-\beta^*_k &=& - \tilde{\alpha}'_k \tilde{\beta}^*_k +
\tilde{\beta}'^*_k \tilde{\alpha}_k \; .
\end{eqnarray}
As a check, note that when $| BD \ket = | BD' \ket$ the Bogolubov
coefficients $\alpha_k$ and $\beta_k$ become trivial. Once the
Bogolubov coefficients $\tilde{\alpha}'_k$, $\tilde{\beta}'_k$,
$\tilde{\alpha}_k$ and $\tilde{\beta}_k$ are known, these expressions
allow us to immediately determine the Bogolubov coefficients relating
the $| BD \ket$ the $| BD' \ket$ state.

\subsection{Relating the two in-states}

The Bogolubov transformation guarantees that both the $| BD
\ket$ and the $|BD'\ket$ in-state can be realized as squeezed
out-states:
\begin{equation}
| BD \ket = \tilde{S} \, | {\rm out}\ket \quad , \quad | BD' \ket =
\tilde{S}' \, | {\rm out} \ket \; ,
\end{equation}
where $\tilde{S}$ and $\tilde{S}'$ are unitary squeezing operators:
\begin{equation}
\tilde{S}= \prod_k \left(
1-|\tilde{\gamma}_k|^2\right)^{\tfrac{1}{4}} \, \exp{\left(
\tfrac{1}{2} \tilde{\gamma}_k b^\dagger_k b^\dagger_{-k} \right)} \; .
\end{equation}
The other squeezing operator $\tilde{S}'$ is obtained by replacing
$\tilde{\gamma}_k$ with $\tilde{\gamma}'_k$. Demanding (\ref{1c})
and expressing the annihilation operators for $|BD \ket$ and $|BD'
\ket$ in terms of $b_k$ and $b^\dagger_k$, we find
\begin{equation}
\tilde{\gamma}_k \equiv -\frac{\tilde{\beta}^*_k}{\tilde{\alpha}_k}
\quad , \quad \tilde{\gamma}'_k \equiv -
\frac{\tilde{\beta}'^*_k}{\tilde{\alpha}'_k} \; .
\end{equation}
This is easiest to show by writing $b_k$ as $\frac{d}{d\,
b^\dagger_k}$ in the Bogolubov transformations. We can also show
(see for instance \cite{Mot,HKP}) that
\begin{eqnarray}
\tilde{\gamma}_k &=& \frac{\bra {\rm out} | \, b_k \, b_{-k} \, | {\rm
BD} \ket}{\bra {\rm out} | {\rm BD} \ket} \nonumber \\
\tilde{\gamma}'_k &=& \frac{\bra {\rm out} | \, b_k \, b_{-k} \, |
{\rm BD'} \ket}{\bra {\rm out} | {\rm BD'} \ket} \; .
\end{eqnarray}
In words: the two $\tilde{\gamma}_k$'s are the probability amplitudes for
pair-creation in the Bunch-Davies and modified Bunch-Davies state.

Now recall that we defined the Bunch-Davies and modified
Bunch-Davies state through the absolute value of exactly these
squared expectation values in (\ref{BD}). As a
consequence the absolute values squared of the Bogolubov ratios
$\tilde{\gamma}_k$ and $\tilde{\gamma}'_k$ are
\begin{equation}
\label{gammas} |\tilde{\gamma}_k|^2 = e^{-\beta \omega_k} \quad ,
\quad |\tilde{\gamma}'_k|^2 = e^{\Delta S(\omega_k)} \; .
\end{equation}
Given these expressions for $\tilde{\gamma}_k$ and
$\tilde{\gamma}'_k$ we are now in a position to calculate $\gamma_k
\equiv -\frac{\beta^*_k}{\alpha_k}$, using (\ref{6}). We find
\begin{equation}
\label{gamma} \gamma_k = \tilde{\gamma}_k
\frac{\tilde{\alpha}'_k}{\tilde{\alpha}_k^{\prime *}} \frac{1 -
\tilde{\gamma}'_k/\tilde{\gamma}_k}{1 -
  \tilde{\gamma}_k\tilde{\gamma}_k^{\prime *}} \; .
\end{equation}
Now, multiplying one set of operators, say $\{b_k\}$, by overall
phases clearly does not affect the choice of state. We can use this
rescaling freedom to make $\tilde{\alpha}'_k$ real. (Indeed, by
further rescalings, we can also force $\tilde{\alpha}_k$ to be
real.) But to proceed, we also need to know something about the
phase of $\tilde{\gamma}'_k$. In fact, it turns out that neither
$\tilde{\gamma}'_k$ nor $\tilde{\gamma}_k$ have any nontrivial
phases; they are both real and positive.

There are a couple of ways to see this. Generally, probability
amplitudes for tunneling -- which is what the $\tilde{\gamma}_k$'s
represent -- are real. For example, in nonrelativistic quantum
mechanics, the probability amplitude for a particle to tunnel across a
potential barrier $V(x)$ is
\begin{equation}
\frac{\psi(x_{II})}{\psi(x_{I})} = \exp \lf i \int_{x_I}^{x_{II}} \! p ~dx
\rt \; ,
\end{equation}
where the momentum, $p$, is purely imaginary: $p =
\sqrt{2m(E-V(x))}$. Thus there is no nontrivial phase in the
amplitude.

Moreover, one can show that if $\tilde{\gamma}_k$ is real and
positive, then $\tilde{\gamma}'_k$ must be real and positive as
well. That is, if the amplitude for tunneling without back-reaction
has no phase, then neither does the amplitude when back-reaction is
incorporated. To see this, imagine the particle or shell of energy
$\omega_k$ being made up of a large number, $N$, of smaller
noninteracting sub-shells, each of energy $\omega_k/N$. As $N$
becomes infinite, the constituent shells have negligible energy and
Hawking's thermal formula, which neglected back-reaction, becomes
exact. The probability amplitude for emitting one of these
sub-shells is then $\exp(-\frac{1}{2}\beta \omega_k/N)$,
where $\beta$ is the inverse de Sitter temperature. But now the probability
amplitude for the finite shell is just the product of the
probability amplitudes for the sub-shells, taking into account that
$\beta$ changes infinitesimally with each emission. Hence
\begin{equation}
\tilde{\gamma}'_k = \prod^N e^{-\frac{1}{2} \beta \omega_k/N} =
e^{-\frac{1}{2}\sum \beta \omega_k  /N} \to e^{-\frac{1}{2}\int
\beta
  d \omega} \; .
\end{equation}
Writing $d M = - d \omega_k$ and invoking the first law of
thermodynamics, $\beta d M = dS$, we obtain
\begin{equation}
\tilde{\gamma}'_k = e^{\frac{1}{2}\Delta S (\omega_k)} \; .
\end{equation}
This is perhaps the easiest way to see that the tunneling
probability had to take the form $\exp(\Delta S)$. But in addition
one learns that, since the infinitesimal probability amplitudes
had no phase, $\tilde{\gamma}'_k$ is real and positive. As a final
check, we confirm explicitly that these arguments are borne out by
considering the particle spectrum of the $\alpha$-states, for which
the result is known. This is done in the appendix. We indeed find
that $\tilde{\gamma}_k$ has no nontrivial phase. Then, by the above
arguments, neither does $\tilde{\gamma}'_k$.

Thus we can write (\ref{gamma}) as
\begin{equation}
\label{abs-gamma} \gamma_k = \frac{ \tilde{\gamma}_k -
\tilde{\gamma}'_k} { 1- \tilde {\gamma}_k \tilde{\gamma}'_k } \; .
\end{equation}
Plugging in (\ref{gammas}) yields
\begin{equation}
\label{10a}
\gamma_k = \frac{e^{-\tfrac{1}{2} \beta \omega_k} -
  e^{\tfrac{1}{2}\Delta S(\omega_k)}}{1- e^{-\tfrac{1}{2}\beta
  \omega_k} e^{\tfrac{1}{2}\Delta S(\omega_k)}} \; .
\end{equation}
As a check, we note that $\gamma_k$ vanishes in the limit $\Delta
S(\omega_k) = - \beta \omega_k$, i.e. the corrections due to
back-reaction vanish when $\frac{\omega_k}{M_p} \rightarrow 0$.

In (\ref{lowenergy}) we found that
\begin{equation}
\label{11a} \Delta S(\omega_k) \approx - \beta \omega_k \lf 1+
\frac{\omega_k H}{8 \pi M_p^2}\rt \; .
\end{equation}
Expanding the exponential in (\ref{10a}) to lowest order in
$\omega_k/M_p$, and using $\beta= 2\pi/H$, we find that
\begin{equation} \label{11b}
\gamma_k \approx
\frac{e^{\pi\frac{\omega_k}{H}}} {e^{2\pi \frac{\omega_k}{H}}-1 }
~ \frac{\omega_k H}{8 M_p^2} \; .
\end{equation}
The energy $\omega_k$ corresponds to the physical energy of
the emitted quantum as measured by a static observer at $r=0$. In
terms of comoving momentum, which is the natural momentum label in
planar coordinates, the physical energy $\omega_k$ is related to
comoving momentum by the scale factor: $\omega_k \propto
\frac{k}{a(t)}$. We will need this relation in the next section, when
we calculate the primordial spectrum of inflationary perturbations
using the modified Bunch-Davies state.

\section{Primordial Spectrum of Inflationary Perturbations}

We have argued that the standard Bunch-Davies vacuum state needs to be
modified because it is inconsistent with energy conservation. Rather,
the spectrum of perturbations should be calculated instead in the
modified state $| BD' \ket$. Now, the standard primordial power
spectrum of inflationary perturbations is directly proportional to the
massless scalar field spectrum; the results we obtain for the scalar
field therefore also pertain to the spectrum of inflationary density
perturbations, as is usual in such calculations.

Consider then an arbitrary vacuum state $| X \ket$. Let $|X \ket$ be
annihilated by all the annihilation operators $a'_k$, and call the
corresponding Fourier-transformed mode functions $u'_k(\eta)$.
The equal-time variance of the scalar field in $| X \ket$,
\begin{equation}
\bra X | \phi(\vec{x}) \phi(\vec{y}) | X \ket = \int \frac{d^3 k}{(2
  \pi)^3} \, |u'_k|^2 \, e^{i \vec{k} (\vec{x} - \vec{y})} \; ,
\end{equation}
defines the power spectrum, $\bra X | \phi^2 |X \ket \equiv \int d\ln
k \, P(k)$, of scalar field perturbations:
\begin{equation}
\frac{2\pi^2}{k^3}\,P(k) = \int d^3 x \, e^{-i\vec{k} \cdot \vec{x}}
\, \bra X | \phi(\vec{x}) \phi(0) | X \ket = |u'_k|^2 \; .
\end{equation}
The well-known scale-invariant result arises when the variance is
computed in the Bunch-Davies state: $|X \ket = |BD \ket$. We denote
this by $P_{BD} \equiv \frac{k^3}{2\pi^2} \, |u_k|^2$. Now let $|X
\ket = |BD' \ket$, and call $P(k)$ the power spectrum in the modified
state. From the Bogolubov transformation
\begin{equation}
u'_k(\eta) = \alpha_k u_k(\eta) + \beta_k u_k^*(\eta) \; ,
\end{equation}
it is straightforward to find the relation between the two power
spectra. We know that we can set $\alpha_k$ to be real. Moreover, in
the previous section we showed that $\gamma_k \equiv -
\beta^*_k/\alpha_k$ is real. Also, write $u_k = |u_k| e^{i\delta}$. Then
we conclude that
\begin{equation}
\label{P-rel} P(k) = \frac{1}{1-\gamma_k^2} \lf
  1-2\gamma_k \cos(2\delta) + \gamma_k^2 \rt ~
  P_{BD} \; .
\end{equation}
When calculating the power spectrum of inflationary perturbations one
is usually instructed to evaluate the spectrum for different $k$ at
the time of horizon crossing, corresponding to $k = a(t) H$. The
reason this gives the right answer is that, after horizon crossing,
the mode-function very quickly approaches a constant. Furthermore,
the Bunch-Davis mode-functions also have the property that the phase
$\delta$ is fixed and independent of $k$, after imposing $k=aH$.
We can determine $\delta$ from the expression for the Bunch-Davies modes,
(\ref{deSittermodes}). Also, since $\gamma_k \ll 1$, we can write
\begin{equation}
\label{P-approx} P(k) \approx \left. \lf 1- 2 \gamma_k
\cos(2 \delta) \rt \right|_{k=aH} ~ P_{BD} \; .
\end{equation}
Now we simply plug in the expression for $\gamma_k$, (\ref{11b}), to
find the leading correction to the primordial power spectrum from
picking the modified Bunch-Davies vacuum $|BD' \ket$. We also have to
impose the condition $k=aH$ (which leads to $\cos(2 \delta) \approx -1$).
As noted above, this condition tells us
to evaluate the expression for the power spectrum at the time the
physical momentum equals the Hubble scale $H$. Since the expression
for $\gamma_k$ is in terms of the physical energy scale, $\omega_k$,
we must set $\omega_k = H$. Thus we find
\begin{equation}
\label{result}
P(k) \approx P_{BD} \left[ 1+\frac{1}{4e^{\pi}} \,
\lf \frac{H(k)}{M_p} \rt^{\! 2} \right] \; .
\end{equation}
This is our final result. We have introduced a $k$-dependence because,
in a generic slow-roll inflationary model, $H$ depends on $k$. The
deviation from unity of the term in brackets represents the
contribution of the most universal quantum gravity effect --
back-reaction due to energy conservation -- to the primordial
power spectrum of inflationary perturbations.

\section{Discussion}

Our calculation shows that the inclusion of back-reaction
results in a mild breaking of de Sitter invariance with a
corresponding mild effect on the CMB power spectrum. This can be seen
by noting that the Bogolubov coefficients relating the modified
Bunch-Davies state to the standard de Sitter invariant vacuum depend
on the momentum scale $\omega_k$. Because the set of all de Sitter
invariant states is paramerized by $k$-independent Bogolubov
coefficients (the $\alpha$-vacua) this necessarily implies that the
modified Bunch-Davies state breaks de Sitter invariance. Moreover, as
the conservation of energy is mandatory, this observation could
perhaps have other important consequences, such as for the stability
of de Sitter space.

The correction we computed can intuitively be understood as coming
from a slightly modified effective de Sitter temperature. By instead
regarding the effect as a modification of the Bunch-Davies vacuum we
were able to compute the exact coefficient of the
correction. Generically, modifications to the initial state
during inflation, as studied in \cite{EGKS3, Dan1, BEFT1},
give rise to a quite specific  modification to the power spectrum: an
oscillatory correction with amplitude on the order of
$(H/M)$ where $M$ is the scale at which the modification
is implemented. The fact that the power spectrum correction
we compute here is not of that form, but is instead a shift
of order $(H^2/M_p^2)$ -- exactly what one expects
from including higher order operators in a bulk effective
field theory \cite{KKLSS} -- suggests
that there may be a bulk effective field theory description
of the modification we've studied. In particular, a correction of this
magnitude generically arises from a Planck-scale-suppressed dimension $6$
irrelevant operator; it would be interesting to study the exact
details and the interpretation in effective field theory of the
operator responsible for this gravitational correction\footnote{In principle, 
this also implies that one-loop quantum corrections to the power spectrum 
can conceivably be of the same order as the correction we find.}.

Our final formula, (\ref{result}), obviously implies a tiny correction,
already highly constrained by $H/M_p \leq 10^{-5}$. Nevertheless, the
result is interesting for a couple of reasons. First, it is
unavoidable, originating as it does in energy conservation. And
second, the correction can be calculated explicitly without requiring
any knowledge of new Planck scale physics. In that sense, it is a
universal imprint of quantum gravity in the large-scale structure of
the universe.

\acknowledgments

We thank Daniel Kabat for several valuable discussions. In addition we would 
like to acknowledge Richard Woodard and Steven Weinberg for valuable correspondence. 
The authors are supported in part by DOE grant DE-FG02-92ER40699. 
M.~P. is a Columbia University Frontiers of Science Fellow.

\appendix

\section{Particle Spectrum in the $\alpha$-States}

Rather than choosing the Bunch-Davies state, which gives rise to a
thermal spectrum for a free-falling timelike observer, one could just
as well have picked one of the de Sitter-invariant $\alpha$-states instead.
As an illustration of our general Bogolubov method we shall calculate here
the spectrum detected by an inertial observer in an $\alpha$-state.
We will reproduce the known result
\cite{Burgess, BMS, KKLSS}, a result usually obtained by carefully
analyzing the singularity structure of the Wightman Green's function.

Now, by definition, the operators, ${a'}_k$, that annihilate an
$\alpha$-state are obtained from the Bunch-Davies operators, $a_k$,
by a Bogolubov transformation,
\begin{equation} \label{12a}
{a'}_k = \alpha^*_k ~ a_k - \beta^*_k ~ a^\dagger_{-k} \; ,
\end{equation}
where the Bogolubov coefficients are $k$-independent:
\begin{eqnarray} \label{12b}
\alpha_k &=& \frac{1}{\sqrt{1-e^{\alpha + \alpha^*}}} \nonumber \\
\beta_k &=& \frac{e^{\alpha}}{\sqrt{1-e^{\alpha + \alpha^*}}} \; .
\end{eqnarray}
Here $\alpha$ is a complex number, with ${\rm Re}(\alpha)<0$, that
labels the $\alpha$-state (and is not to be confused with the
Bogolubov coefficient $\alpha_k$); ${\rm Re}(\alpha) = - \infty$
corresponds to the Bunch-Davies state.

The operators, $b_k$, that annihilate the out-state are
themselves related by a Bogolubov transformation to the Bunch-Davies
creation and annihilation operators:
\begin{equation} \label{13a}
b_k = \tilde{\alpha}^*_k ~ a_k - \tilde{\beta}^*_k ~ a_{-k}^\dagger \; .
\end{equation}
Now, multiplying $b_k$ by an overall phase obviously does not affect
the choice of vacuum state. Use this freedom to fix the phase so that
$\tilde{\alpha}_k$ is real. Next, express these operators in terms of
the $\alpha$-state operators:
\begin{equation} \label{13b}
b_k = \tilde{\alpha}^{\prime *}_k ~ {a'}_k - \tilde{\beta}^{\prime
*}_k ~ {a'}^\dagger_{-k} \; .
\end{equation}
Then, by inverting (\ref{12a}),
\begin{equation} \label{13c}
a_k = \alpha_k ~ {a'}_k + \beta^*_k ~ {a'}_{-k}^\dagger \; ,
\end{equation}
and substituting into (\ref{13a}), we conclude that
\begin{eqnarray} \label{14}
\tilde{\alpha}^{\prime *}_k &=& \tilde{\alpha}^*_k \alpha_k -
\tilde{\beta}^*_k \beta_k  \nonumber \\
\tilde{\beta}^{\prime *}_k &=& \tilde{\beta}^*_k \alpha^*_k -
\tilde{\alpha}^*_k \beta^*_k \; .
\end{eqnarray}
Or, using (\ref{12b}),
\begin{eqnarray} \label{15a}
\tilde{\alpha}^{\prime *}_k & = & \frac{1}{\sqrt{1-e^{\alpha +
\alpha^*}}} \left[ \tilde{\alpha}^*_k - e^{\alpha}
\tilde{\beta}^*_k \right] \nonumber \\
\tilde{\beta}^{\prime *}_k & = & \frac{1}{\sqrt{1-e^{\alpha +
\alpha^*}}} \left[ \tilde{\beta}^*_k - e^{\alpha^*}
\tilde{\alpha}^*_k \right] \; .
\end{eqnarray}
The spectrum is controlled by the absolute value of the parameter
$\tilde{\gamma}'_k \equiv -\frac{\tilde{\beta}^{\prime
*}_k}{\tilde{\alpha}'_k}$. Now we use the fact that, for the thermal
Bunch-Davies state, $|\tilde{\gamma}_k|^2 = e^{-\beta \omega_k}$.
Then a little algebra yields
\begin{equation} \label{16a}
| \tilde{\gamma}'_k |^2 = e^{-\beta \omega_k} \left |
\frac{1+e^{\alpha^*} \tilde{\gamma}_k^{-1}}{1+e^{\alpha^*}
  \tilde{\gamma}_k^*} \right|^2 \; .
\end{equation}
In section 4.2, we argued that, on general grounds, $\tilde{\gamma}_k$
is purely real. This allows us to write $\tilde{\gamma}_k =
\exp{(-\tfrac{1}{2} \beta \omega_k)}$, and we obtain
\begin{equation} \label{16b}
| \tilde{\gamma}'_k |^2 = e^{-\beta \omega_k} \left |
\frac{1+e^{\alpha + \tfrac{1}{2} \beta \omega_k}}{1+e^{\alpha -
\tfrac{1}{2} \beta \omega_k}} \right|^2 \; ,
\end{equation}
in perfect agreement with previously derived results
\cite{BMS}. (Had we not known that $\tilde{\gamma}_k$ was purely real,
we would have learned that by matching our result with that in the
literature.) Note incidentally that, for ${\rm Re}(\alpha) \neq -
\infty$, the spectrum is not thermal; in the limit $\frac{\omega_k}{H}
\rightarrow \infty$, for instance, we find $|\bar{\gamma}_k|^2 =
e^{\alpha + \alpha^*}$, corresponding to constant occupation numbers.

\end{document}